\documentclass[12pt]{article}

\def\demi{{\textstyle {1\over2}}}

\def\nn{\nonu}
\def\bea{\begin{eqnarray}}
\def\eea{\end{eqnarray}}
\def\be{\begin{equation}}
\def\ee{\end{equation}}

\def\Q{s_n }
\def\w{\wedge}

\def\t{\tilde}

\def  \Tr{ {\bf\rm Tr}  }
\def\m{{\bar m}}
\def\n{{\bar n}}
\def\p{{\bar p}}
\def\q{{\bar q}}
\def\l{{\bar l}}\let\nonu=\nonumber
\def\M{{\bar M}}
\def\N{{\bar N}}
\def\P{{\bar P}}
\def\Q{{\bar Q}}

\font\mybb=msbm10 at 12pt

\def\bb#1{\hbox{\mybb#1}}

\def\complex{{\bb{C}}}

\def\real{{\bb{R}}}

\def\A{{\cal A}}

\begin{document}
\bibliographystyle{perso}
\begin{titlepage}
\null 

\hfill PAR--LPTHE P04--32

\hfill SISSA--94/2004/FM

\vskip .5truecm
\begin{center}

{\Large {\bf
Topological Symmetry of forms, \\ 
${\cal N}=1$ Supersymmetry and S--duality on Special Manifolds  
}}
\vskip 6mm
{ {\bf Laurent Baulieu} \\  
Laboratoire de Physique Th\'eorique et Hautes Energies, 
Universit\'e  de  Paris~VI~and ~Paris~VII, France$^{ \dag}$
\\{\bf Alessandro Tanzini}\\ 
S.I.S.S.A., via Beirut 2/4, 34014 Trieste, Italy$^{\dag\dag}$   } 
\end{center}

\vskip 13mm
\noindent
{  Abstract}: 
We study the quantization of a 
holomorphic two--form coupled to a Yang--Mills field on
special manifolds in various dimensions, and we show that it   
yields twisted supersymmetric theories.
The construction determines ATQFT's 
(Almost Topological Quantum Field Theories), that is, 
theories  with  observables  that are invariant under changes of metrics   
belonging  to restricted classes.   
For K\"ahler manifolds in four dimensions,
our topological model is related to ${\cal N}=1$ Super Yang--Mills theory.
Extended supersymmetries are recovered
by considering the coupling with chiral multiplets.
We also analyse 
Calabi--Yau manifolds in six and eight dimensions, and
seven dimensional $G_2$ manifolds of the kind recently discussed by Hitchin.
We argue that the two--form field could play an interesting r\^ole for the
study of the conjectured S--duality in topological string.
We finally show that in the case of real forms 
in six dimensions the partition
function of our topological model is related
to the square of that of the holomorphic Chern--Simons theory,
and we discuss the uplift to seven dimensions and
its relation with the recent proposals for the topological M--theory.
 
\vfill 
\hrule 
\medskip
\obeylines
$^{\dag}$Postal address: Laboratoire de Physique Th\'eorique et des Hautes 
Energies, Unit\'e Mixte de Recherche CNRS 7589, Universit\'e Pierre et 
Marie Curie, bo\^\i te postale 126.4, pl. Jussieu, F--75252 PARIS Cedex 05. 
e-mail: {\tt baulieu@lpthe.jussieu.fr}

$^{\dag\dag}$e-mail: {\tt tanzini@sissa.it}

\end{titlepage}

\section{Introduction}

The idea that  Poincar\'e supersymmetry is a "phase" of a more fundamental symmetry is appealing.  In a series of earlier works, using various examples, it was shown  that  Poincar\'e supersymmetry and  topological symmetry are deeply related. 
We have shown in \cite{sugra11} that the field spectrum of dimensionally reduced 
${\cal N}=1$ $ D=11$  supergravity  can be determined  in the context of an 8-dimensional gravitational Topological Quantum Field Theory (TQFT). More precisely, the equivalence of the supergravity and topological actions was  shown up to quartic fermionic terms, around a $Spin(7)$ invariant vacuum.
One foresees from this result that many models, which are dimensional reductions and truncations of maximal supergravity, might be possibly related by twist to topological models \cite{4d}. 
The BRST operator that characterizes  a topological   symmetry is a scalar operator which  can be defined in any given curved space, while Poincar\'e supersymmetry  is a delicate concept in curved space. Therefore, topological symmetry could be a more fundamental concept than  Poincar\'e supersymmetry. On the other hand, in order to perform the twist operation that relates Poincar\'e supersymmetry and  topological symmetry, one often needs to use  manifolds with special holonomy.
A possible derivation of supergravity from the TQFT of a 3-form in higher dimensions
was discussed in \cite{delire}. 

On the other hand, the relevance of the 8-dimensional topological Yang--Mills theory
\footnote{This theory can  untwisted and ``dimensionally oxidated'' in the $N=1, d=10$ Yang--Mills theory \cite{bakasi}.} and its coupling to a 3-form, and of its dimensional reductions in lower dimensions, was emphasized for the construction of M/F theory in \cite{bakasi, acharya, blau}. Recently,
these theories received a renewed interest  in the context of topological
string theory \cite{iqbal,ooguri,kapustin,strominger} 
and its possible generalizations
to M theory \cite{Dijgk,samson,grassi}. 

In this paper, we will discuss the quantization of (holomorphic) two--forms
coupled to a Yang-Mills field on special manifolds in various dimensions.
These theories are basically ATQFT's (Almost Topological Quantum Field Theories),
in the sense that they are defined in terms of a classical action and a set of
observables which are invariant under changes of coordinates belonging to restricted classes, for instance, reparametrizations that 
respect a complex structure.  This is to be compared to genuine TQFT that contain observables invariant under all possible changes of metrics. Interesting cases 
that we will analyse in detail are K\"ahler manifolds in four dimensions and 
special manifolds in higher (6,7 and 8) dimensions. In particular in seven dimensions
we will analyse $G_2$ manifolds of the kind recently studied by Hitchin \cite{Hitchin}. 

One of the original motivations for this work was to try to understand 
how (twisted) $N=1$ supersymmetric theories can be directly constructed as TQFT. 
As we will see in the next section, this immediately leads to the introduction
in the classical action of a ``charged" 2-form $B$, valued in the adjoint
representation of a Lie algebra. In these models one can also 
consider the coupling to chiral multiplets. If these transforms
in the adjoint representation, one recover in this way also the extended 
supersymmetry in a twisted form. 

It was noticed that 
the presence of a two--form field helps in clarifying the S--duality properties of
Yang--Mills theory in four dimensions \cite{bfym} and of its supersymmetric
extensions in various dimensions \cite{basha}.
Something similar seems to happen also in the context of the 
topological string. In fact in the six--dimensional case we will see that by 
choosing different
gauge--fixing conditions on the $B$ field one can get a theory
related to the B model (more precisely, to the holomorphic Chern--Simons) 
or to the A model, as a twisted maximally supersymmetric Yang--Mills.
Thus one can expect that the study of this model can give interesting 
informations about the S--duality of topological string 
\cite{neitzke,ooguri,kapustin}
and on the relation between Gromov--Witten and Donaldson--Thomas
invariants conjectured in \cite{panda}.
  
In the last part of the paper we will consider the case of real forms. 
We will show that the partition function
of our topological model is six dimensions is formally equal to the square
of the partition function of the holomorphic Chern--Simons theory.
This indicates that this topological BF model could play some r\^ole
in clarifying the relationship between the black hole 
and the topological string partition functions pointed out
in \cite{strominger}. As a further remark in this direction, we observe that 
the uplift of our BF model to some particular seven dimensional $G_2$ Hitchin's
manifolds is related to the 
topological M-theory recently discussed by Dijkgraaf et al. \cite{Dijgk}.
We remind that BF models in three and four dimensions
are directly related to gravitational theories \cite{Horowitz}\footnote{It is 
interesting to observe that in four dimensions these models describe 
perturbations around gravitational instantons.
For vanishing cosmological constant, these solutions are also
described by the twisted
supergravity models discussed in \cite{4d}.}.
It is then natural to investigate whether a similar relationship
can be found on more general grounds
for the higher dimensional models that we study in this paper,
by performing a suitable choice of the gauge group 
and of the gauge--fixing conditions \cite{w}.
These relations are the generalisation to higher
dimensions of the description of three--dimensional
gravity with a Chern--Simons model \cite{witten3d}.

The organization of the paper is as follows: in Sect.2
we introduce the holomorphic BF model and 
discuss its relationship with ${N}=1$
(twisted) supersymmetry.
Notice that the quantization of this model requires 
the use of the Batalin--Vilkoviski formalism.
In Sect.3 the four
dimensional case is considered, including a detailed discussion on the
coupling with a chiral multiplet. In Sect.4 we discuss the six--dimensional
case on a Calabi--Yau three--fold and show how two different quantizations
yields respectively to theories related to the B model and A model of topological string.
In Sect.5 we discuss the eight--dimensional theory on 
a Calabi--Yau four--fold and its dimensional reduction to CY$_3\times S^1$.
The eight--dimensional model is discussed also for manifolds
with $SU(4)$ structure. This theory can be regarded as a generalization of the
four--dimensional self--dual Yang--Mills model \cite{siegel}. 
Finally in Sect.6 we consider the real BF model in six 
and seven dimensions and its relationship with the topological M theory.

\section{$N=1$ supersymmetry and the holomorphic BF theory}

The standard construction of a TQFT leads to models with
$N=2$ supersymmetry. To see this, let us consider  
the "prototype" case of Topological Yang--Mills theory
in four and eight dimensions. The relevant BRST transformations read 
\bea\label{fond} 
 \matrix{\delta A _\mu&= &\Psi  _\mu +D_\mu  c
&
 \delta \Psi _\mu&= &D _\mu\Phi  -[ c,\Psi  _\mu]
\cr
  \delta c&= & - \Phi - \demi [c,c]
&
  \delta \Phi &= & - [c,\Phi]}
   \eea
These equations stand for the geometrical identity  $(\delta+d)(A+c)+\demi [A+c,A+c]=F+\Psi+\Phi$ \cite{basi}. There are as many components in the 
topological ghosts as  in the gauge fields, and to gauge fix the topological freedom, one must also introduce as many antighosts as topological ghosts. 
The antighosts are    an anticommuting antiself dual 2-form  $\kappa_{\mu\nu} $ and an anticommuting  scalar $\eta$. For each one of the antighosts, there is an associated  Lagrange multiplier field, and their BRST equations are : 
  \bea\label{fond'} 
 \matrix {
 \delta\kappa_{\mu\nu} &=&b_{\mu\nu}  -[ c,\kappa_{\mu\nu} ]
&
  \delta b_{\mu\nu} &=&- [c, b_{\mu\nu} ]
   \cr
 \delta\bar  \Phi &=&\eta - [c,\bar \Phi]
 &
    \delta   \eta &=& [c,  \eta]
    \cr
    }
   \eea
The twist operation is a mapping from  these ghost and antighost  fermionic degrees of freedom on a pair of spinors, which leads one to reconstruct the spinor spectrum of $N=2$ supersymmetry,  both  in 4 and 8 dimensions.      
The scalar BRST operator $\delta$ can then be identified  as a Lorentz scalar combination of the $N=2$  Poincar\'e supersymmetry generators. However, this ''twist" 
operation has different geometrical interpretation in 4 and 8 dimensions. In the former case, it is a redefinition of the Euclidean Lorentz group contained in the global    $SU_L(2)\times SU_R(2)\times SU(2)$ invariance  of the supersymmetric theory. In the latter case, it uses the triality  of 8-dimensional space. In the previous works
\cite{bakasi,acharya,blau}, a constant covariant spinor has been used, which implies that one uses $Spin(7)$ invariant manifolds; one can also use a manifold with $SU(4)$ holonomy.

Using self-duality equations as gauge functions, one can build a $\delta$-exact action that   provides twisted supersymmetric theories with a $\delta$-exact energy momentum tensor. The   cohomology of the $\delta$ symmetry determines  therefore  a ring of topological observables, which is a subsector of
the familiar set of observables for the   
gauge particles. The latter  is selected from the cohomology of the  ordinary gauge invariance.

From (\ref{fond}) and (\ref{fond'}) one concludes that in these TQFT one has
twice as many fermionic degrees of freedom than bosonic ones. 
This makes seemingly impossible to determine   
${N}=1$ models as a twist of TQFTs.  
We may, however,  look for models with a ``milder" topological symmetry, such as BF models or Chern--Simons type model, characterized by metric independent classical actions.
Such actions are not boundary terms, and thus their  topological symmetry   cannot be as large as that displayed in (\ref{fond}) and   (\ref{fond'}). This might lead us to models that are twisted ${N}=1$ supersymmetric theories.
In this paper, we will consider the following holomorphic BF action
\be
I_{n-BF} = \int_{M_{2n}} \Tr (B_{n,n-2}\w F_{0,2})
\label{hBF}
\ee
which is defined on any complex manifold $M$ of complex dimension
$n$. Some aspects of the classical action (\ref{hBF}) 
were studied in \cite{popov}. 
It would be interesting
to study its quantization and possible relation with supersymmetry.
For the moment we only consider some particular models which can be obtained
from (\ref{hBF}) by choosing a particular form for the field $B_{n,n-2}$.
Notice that the equations of motion for this field coming from
(\ref{hBF}) implies that $B_{n,n-2}$ is a holomorphic $(n,n-2)$ form.

One might  try to define a theory that is classically invariant under the 
following (almost) topological symmetry, which is localised in the 
holomorphic sector
 \footnote{We use the standard notation  where the    
   complex indices are denoted with latin letters $m,n$ and $\m,\n$, and 
   the complex coordinates are given by $z^m$ and $\bar z^{\bar m}$.}:
 \bea\label{hol} 
 \matrix {Q A _m= \Psi_m +D_m  c 
  &
 Q \Psi _m = -[ c,\Psi  _m]
\cr
  Qc=- \demi [c,c]
&
  Q A_\m =D_\m c}
   \eea
This ``heterotic" symmetry was  already  used 
\cite{Johansen, Witten1, Park, Dijkgraaf1} in four dimensions and in 
\cite{bakasi, Park-cy3} in higher dimensions.
Here we recover it as a symmetry associated to the classical action (\ref{hBF}).
If we count the ghost degrees of freedom, we have 2 
components for $\Psi_m$ 
and one for $c$.  
Notice that $\Psi_m$  cannot have a  ghost of ghost symmetry  
with a ghost of ghost $\Phi$,  since  
$  Q\Psi_m =D_m\Phi -  [c,\Psi_m]$ and 
 $  Qc=\Phi - \demi [c,c]$ would imply that $Q^2A_\m \neq 0$.
Modulo gauge transformations, only one degree of freedom 
for the field $A$ is left free by the symmetry in (\ref{hol}).
Moreover, if we succeed  in writing  a BRST gauge--fixed  action for the classical symmetry in (\ref{hol}), this will depend on the ghosts $\Psi_m$ and as many   antighost components as there are in  $\Psi_m$ (four components).
Then the number of  fermion degrees of freedom will  fit with those of  a single Majorana spinor
and we   have a chance to eventually  
reach ${N}=1$ supersymmetry, 
as we will explain in detail in the next section.
Notice that in these models one can also recover the coupling to a
chiral multiplet in the adjoint representation and the corresponding
extended supersymmetry in the twisted form.

\section{Four dimensions: K\"ahler manifold} 

 \subsection{The classical action for a $BF$  system on a K\"ahler manifold}

On a K\"ahler manifold one can define a complex structure 
\bea
\label{J}
&&J^{mn} = 0 \ , \quad\quad J^{\m\n}=0 \ , \nn \\
&&J^{m\bar n} = i g^{m\bar n}
\eea
which allows one to introduce complex
coordinates 
$z^m$ and $z^{\bar m}$ and $1\leq m, \m\leq N $ in 2N dimensions   by 
\be
J^m_n z^n = i z^m \ , \quad\quad J^\m_\n z^\n = - i z^\m \ .
\label{cplx}
\ee
In four dimensions, the action (\ref{hBF}) reads
\bea
\label{BF}
I_{cl} (A,B)=\int _{{\it M}_4}  \Tr
B_{2,0}\w
F_{0,2}=
\int _{{\it M}_4} d^4x \sqrt{g} \  \Tr
\Big(\epsilon^ {mn\m\n}
B_{mn} 
F_{\m\n}\Big)
\eea
where $F=dA+A\w A$ is the curvature of the Yang--Mills field $A$.
The equations of motion are 
\bea
F_{\m\n}=0
\quad
\quad
 \epsilon^ {mn\m\n}
 D_\n B_{mn}=0
\eea
Classically, $A_m$ is undetermined, $A_\m$ is a pure gauge  and $B_{mn}$ is holomorphic.
Notice that $B_{2,0}$ has no vector gauge invariance. It counts for one propagating degree of freedom. Altogether, there are two gauge invariant degrees of freedom that are not specified classically.
Modulo gauge invariance, there  is   a  mixed propagation  between $A$ and $B$.
The symmetries of the action (\ref{BF}) are 
\bea\label{holB} 
 \matrix {Q A _m= \Psi_m +D_m  c 
  &
 Q \Psi _m = -[ c,\Psi  _m]
\cr
  Qc=- \demi [c,c]
&
  Q A_\m =D_\m c 
\cr
  Q B_{mn}=  -[ c, B_{mn} ] }
\eea
In the first two lines of the above equation we can recognize the symmetry
(\ref{hol}).
The geometrical interpretation of complete charged 2-forms  is a non-trivial 
issue. However, in the language where form degree and positive and negative 
ghost number are unified within a bigrading, the charged 2-form can be 
understood as a sort of Hodge dual to the 
Yang--Mills field \cite{bara}. Here we only consider the $(2,0)$ component 
of such an object, and we hopefully avoid the ambiguities for defining its 
theory.  
We now explain the  BRST quantization of the action~(\ref{BF}), 
for the sake of inserting it in a path integral.

\subsection{Quantization of  the  $BF$  system on a K\"ahler manifold}
 
In order to define a quantum theory, that is, a path integral,   
we need to gauge fix the (almost) topological symmetry of the $BF$ 
system, in a way that respect the BRST symmetry associated to this symmetry. 
As it is well-known, the anti--self--duality condition in 4 dimensions 
can be expressed in complex coordinates as:
\bea\label{self}
&&F_{mn}=0 \ , \quad \quad  F_{\m\n}  = 0  \nn \\
&&J_ {m\n}  F^{m\n}  = 0 
\eea
and one has the identity 
\bea\label{selfym}
\Tr \left( F_{\m\n}F^{\m\n} + \demi |J_{m\n}  F^{m\n}|^2\right)=
\frac{1}{4}\Tr ( F_{\mu\nu}F^{\mu\nu} + F_{\mu\nu}\t F^{\mu\nu})
\eea
Modulo ordinary gauge invariance,  we have two topological freedoms, 
corresponding to the two components in $\Psi_m$. 
In order to perform a suitable gauge fixing for the two--form $B_{2,0}$ 
and for $A_{1,0}$, which is the part of the gauge connection absent from
the classical action (\ref{BF}), we introduce two anticommuting antighosts 
$ \kappa^{mn}$ and  $\kappa$,  and two Lagrange multipliers
$b^{mn}$ and $b$: 
\bea\label{la} 
 \matrix {Q \kappa^{mn}&= & b^{mn}
&
  Q b^{mn}&= & 0
 \cr
  Q \kappa &= & b
 & 
Q b&= & 0
}
\eea
Since $Q^2=0$ from the beginning, we have  a  first order 
Batalin-Vilkoviski (BV) system.
However, since the treatment of the chiral multiplet in the next Section 
will produce a non-trivial second order BV system, we find convenient to 
introduce right now BV antifields for $A$ and $B$ and their ghosts, 
antighosts and Lagrangian multipliers. 
The upper left notation $^*$ labels antifields\footnote{As stressed 
in \cite{bara}, 
the antifields of $A$ appear in the ghost expansion of $B$ and vice-versa.}. 
Let us recall that the antifield $^*\phi$ of a field with ghost number 
$g$ has ghost number $-g-1$ and opposite statistics. 
For a $Q$-invariant BV action $S$ one has
$Q\phi ={{\partial_l S}\over {\partial ^*\phi}}$
and
$Q^*\phi = - {{\partial_l S}\over {\partial \phi}}$.
The property $Q^2=0$ is equivalent to the master equation 
\bea
\label{bveq}
{{\partial_r S}\over {\partial \phi}}
{{\partial_l S}\over {\partial ^*\phi}}
=0
\eea
where $(\partial_r, \partial_l)$ indicate respectively the derivatives
from the left and from the right.

The following BV action encodes at once the classical action (\ref{BF})
and the definition of its BRST symmetry:
\bea\label{BVYM}
S &=&
\int _{{\it M}_4} d^4x \sqrt{g} \ \Tr
\Bigl( \frac{1}{4}
\epsilon^ {mn\m\n}
B_{mn} 
F_{\m\n}
\nn\\
&& \ \ \ \
+^*A^{m}   (\Psi_m +D_m  c)
+^*A^{\m}  (D_\m  c)
\nn\\  
&&\ \ \ \
-^*B ^{mn }    [ c, B_{mn} ]
-^*\Psi ^{m}    [ c, \Psi _{m} ]
- \demi  ^*c [c,c]
\nn\\
&&\ \ \ \
+^*\kappa_{mn} b^{mn}
+^*\kappa b
\Bigr)
\eea
The BV master equation (\ref{bveq})
is satisfied, which implies the gauge invariance of the 
classical action as well as the nilpotency $Q^2=0$ on all the fields.
It is actually important to note that the invariance of the action 
(\ref{BVYM}) implies that
\bea\label{bf}
Q^*B^{mn}=
\epsilon^ {mn\m\n} F_{\m\n} - [c,^*B^{mn}] .
\eea
This equation will shortly play a key role for defining the coupling to scalar fields.

The topological gauge fixing corresponds to the elimination of antifields 
by a suitable choice of a gauge function $Z$. 
The antifields  are to be replaced   in the path integral by 
the BV formula: 
\bea \label{bfZ}
^*\phi
=
{{\delta Z}\over {\delta \phi }}
\eea
The $Q$-invariant observables are formally independent on the choice of $Z$. 
In particular, their mean values are expected to be independent on small 
changes of the metric that one must introduce to define $Z$.

In order to concentrate  path integral around the anti-self-duality 
condition (\ref{self}) we choose:
\bea\label{Z}
Z
=
\kappa^{mn}(B_{mn} - \epsilon_ {mn\m\n} F^{\m\n})
+\kappa (\demi b + i J_ {m\n}  F^{m\n} )
\eea
The BV equation (\ref{bfZ}) implies that $B_{2,0}$ is 
eliminated in the path integral with
$B_{mn} = \epsilon_ {mn\m\n} F^{\m\n}$. 
After Gaussian integration on $b$, the gauge-fixed action reads
\bea\label{SYMn=1}
 S^{g.f.} &=&
\int _{{\it M}_4} d^4x \sqrt{g} \  \Tr
\Bigl (
 F_{\m\n}F^{\m\n} + \demi |J_ {m\n}  F^{m\n}  |^2 
\nn \\
&&\ \ \ \ 
-2 \epsilon^{\m\n pq}\kappa_{\m\n}D_{p}\Psi_{q}
+ i
\kappa J^{\m l}D_{\m}\Psi^{l} \Bigr )
\eea
Here and in the following discussions we omit the  
$c$--dependent terms in the action. In fact these terms express the 
covariance of the gauge--fixing conditions (\ref{self}) with respect to 
the gauge symmetry, and vanish when these conditions are enforced.  
Moreover, we leave aside the standard gauge fixing of the 
ordinary gauge invariance $\partial^\mu A_\mu=0$.

The action (\ref{SYMn=1}) can be compared with that of $N=1$ SYM 
on a K\"ahler manifold.
It is known \cite{lawson}
that on a complex spin manifold
the complex spinors can be identified with forms 
$S_{\pm}\otimes \complex \sim \Omega^{0,^{even}_{odd}}$, 
so that we can identify
our topological ghost $\Psi_m$ as a left--handed Weyl 
spinor $\lambda_\alpha$ and the topological anti--ghosts 
$(\kappa_{\m\n},\kappa)$ as a right--handed Weyl spinor 
$\bar\lambda^{\dot\alpha}$. More explicitly,  
the holonomy group of a four--dimensional K\"ahler manifold is locally given by
$U(2) \sim SU(2)_L\times U(1)_R \subset SU(2)_L \times SU(2)_R$, 
so that one can naturally identify the forms $\sigma_{\mu~\alpha\dot 1}dx^\mu$
and $\sigma_{\mu~\alpha\dot 2}dx^\mu$ as 
$(1,0)$ and $(0,1)$ forms respectively \cite{Johansen,Witten1}. 
Then the twist reads\footnote{We 
define the euclidean 
$\sigma$ matrices as $\sigma_\mu=(i\tau^c,{\bf 1})$, $\tau^c, \ c=1,2,3$
being the Pauli matrices.}
\bea\label{twistn1}
\Psi_m &=& \lambda^\alpha\sigma_{\mu \ \alpha\dot 1} e^\mu_m \ , \nn\\
\kappa_{\m\n} &=& 
\bar\lambda_{\dot\alpha}\ \bar\sigma_{\mu\nu \ \dot 2}^{\ \dot\alpha} 
 \ e^\mu_\m e^\nu_\n  \ ,
\nn\\
\kappa&=& \delta^{\ \dot\alpha}_{\dot 2}\bar\lambda_{\dot\alpha}
\eea
On a Hyperk\"ahler manifold, the twist formula can be reinterpreted by making explicit a constant spinor dependence in (\ref{twistn1}). 
With this change of variables, it is immediate to recognize that the action 
(\ref{SYMn=1}) is the $N=1,D=4$   Yang--Mills action
\bea\label{SYMn=1t}
 S_{SYM} &=&
\int _{{\it M}_4} d^4x \sqrt{g} \ \frac{1}{4} \Tr
\Bigl ( F_{\mu\nu}F^{\mu\nu} + F_{\mu\nu}\tilde F^{\mu\nu}
+
\bar\lambda \gamma^\mu D_\mu \lambda \Bigr )
\eea
As compared to \cite{Johansen}, we started from  
a classical BF system, which, eventually,  gives the 
${N}=1,D=4$   Yang--Mills theory  as a microscopic theory in a 
twisted form.   
Let us notice that it should be possible to cast 
the  topological BRST symmetry in the form 
of conditions on curvatures yielding descent equations with asymmetric 
holomorphic decompositions and eventually solve the cocycle equations for $Q$,
similarly to what has been done in \cite{basi}
for the Topological Yang--Mills theory.
  
\subsection{Coupling of the $BF$ to a chiral multiplet}

Since the $N=2$ SYM theory is an ordinary TQFT, and since its Poincar\'e 
supersymmetric version can be obtained by coupling the    
$N=1$   Yang--Mills multiplet to a chiral multiplet in the adjoint 
representation of the gauge group, one expects to have an expression 
of the $N=1$ scalar theory as an (almost)TQFT on a K\"ahler manifold. 
As we shall see, this is slightly more complicated than the $N=1$ Super 
Yang-Mills theory, since it will involve the vector gauge symmetry of a
(0,2)-charged form, and thus a 2nd rank BV system arises.
    
In order to introduce the chiral multiplet, we extend the set of classical 
fields of the previous section as
\bea
B_{2,0} \to (    B_{2,0},    B_{0,2}),
\eea
However, we keep   the same classical action as in (\ref{BF}): 
\be
I_{cl} (A,B_{0,2},B_{0,2})=\int _{{\it M}_4}  \Tr
(B_{2,0}\w
F_{0,2})=\int _{{\it M}_4}  \Tr
(\epsilon^ {mn\m\n}
B_{mn} 
F_{\m\n})
\label{class-chi}
\ee
Having a classical action that is independent of  
$B_{0,2}$ is equivalent of having  the  following symmetry 
for $ B_{0,2}$ 
\bea\label{full}
Q  B_{\m\n} &=& D_{[\m} \Psi_{\n]}
    -[c,   B_{\m\n} ] - \frac{1}{4}
\epsilon_ {\m\n mn} [ ^*B^{mn},\Phi]  \\
Q    \Psi_{\m}
    &=&D_{\m}   \Phi
    -[c,    \Psi_{\m} ] \nn \\
Q      \Phi
    &=& 
    -[c,     \Phi]. \nn
\eea
In fact, the unique degree of freedom carried by $B_{\m\n}$
is canceled by the two degrees of freedom of the topological 
ghost $ \Psi_{\n}$, defined modulo the ghost of ghost symmetry 
generated by $\Phi$.
The presence of the antifield  $^*B^{mn}$ in (\ref{full}) is necessary in 
order that $Q^2=0$, as one can verify by using (\ref{full}), 
and (\ref{bf}), together with the usual BRST variation of the ghost 
$c$, $Qc= -c^2$. 
We thus have a second order BV system, since the BRST variations 
of the fields depend linearly on the antifield.
This non trivial property justifies, a posteriori, that   the classical action depend on the charged 
$(2,0)$-form 
$B_{2,0}$ in an ''almost  topological way", as in (\ref{BF}), with
the ordinary gauge symmetry  $QA_\m= D_\m c$ and $QB_{2,0}=-[c,B_{2,0}]$. 
This determines the relevant $Q$-transformation of the antifield of the 
2-form, which is eventually necessary to obtain a closed symmetry.

The fate of the $(0,2)$-form 
$B_{0,2}$ is to be gauge-fixed and eliminated from the action, as it was the case for 
$B_{2,0}$, but wit a different gauge function.
For this purpose we choose a  topological antighosts that is a 
(0,2)-form $\kappa^{\m\n}$,
with bosonic Lagrange multiplier $b^{\m\n}$. The ghost of ghost symmetry of 
$\Psi_\m$ must   be gauge fixed, and we introduce a bosonic antighost
$\bar\Phi$ with fermionic Lagrange multiplier $\bar\eta$.  $\kappa^{\m\n}$ and 
$\bar\eta$ will be eventually untwisted and provide half of a Majorana spinor for $N=1$ supersymmetry

One has in fact an ordinary pyramidal structure for a 2-form gauge field~\footnote{In this table, we indicate explicitly the ghost number 
of the fields by a superscript. The BRST symmetry acts on the South--West 
direction, as indicated by the arrows.},  
which shows that 
$B_{\m\n}$ truly carries zero degrees of freedom, and can be consistently gauge-fixed to zero
\begin{eqnarray}\label{pyr}
\let\hw=\hidewidth
\matrix{ && &&   &&    &&  B_{\m \n  _{}}       
\cr   &&    &&    &&         &   \hw\swarrow\hw  
\cr   && &&   &&     \Psi^{(1)}_{\m}  &&    &&        \kappa^{(-1) \m\n }  
\cr  && &&   &  \hw\swarrow\hw &&    &&   \hw    \ \ \ \ \ \ \    \swarrow\hw  
\cr  && &&   \Phi^{(2)}  &&    &&      \hw  b^{(0) \m\n } 
\hw &&   &&    \bar\Phi^{(-2)} 
\cr  
&& &     &&&& \hw\hw&&&&  \swarrow\cr  && &&  &&      
\hw &&  &&   \hw \bar\eta^{(-1)} \hw &&      &&      \cr   }%
\end{eqnarray} 
The vector ghost symmetry of the charged $(0,2)$-form  with ghost $\Psi^{(1)}_{\m}$ plays the role of  
a topological symmetry. 
In the untwisted theory,  
$\Phi$  and $\bar\Phi$ will be identified as 
the complex scalar field for the $N=1$ chiral multiplet in 4 dimensions.
The BV action for the fields in the Table (\ref{pyr}) and their antifields  
is
\bea\label{BVHM}
S_{matter} &=&
\int _{{\it M}_4} d^4x \sqrt{g} \ \Tr
\Bigl(
 ^*B^{\m\n}   (D_{[\m} \Psi_{\n]}
    -[c,   B_{\m\n} ] - \frac{1}{4} 
    \epsilon_ {\m\n mn} [ ^*B^{mn},\Phi])
\nn\\  
&&\ \ \ \
+^*\Psi ^{\m }  ( D_{\m}   \Phi
    -[c,    \Psi_{\m} ])
-^*\Phi [ c, \Phi ]
\nn\\
&&\ \ \ \
+^*\kappa_{\m\n} b^{\m\n}
+^*\bar\Phi \bar\eta 
\Bigr )
\eea
For consistency, we have to add to the above action the
action (\ref{BVYM}) in order to properly define the variation of 
$^*B^{mn}$ as in (\ref{bf}).

In order to gauge--fix $S_{matter}$,
we choose the following BV gauge function:
\bea\label{Z'} 
Z'
=
\kappa^{\m\n}B_{\m\n} +\bar\Phi D^\m\Psi_\m
\eea
Using the BV equation (\ref{bfZ}) and integrating 
on the Lagrangian multiplier $b^{\m\n}$ one gets
$B_{\m\n}=0$, and finds
\bea\label{HMn=1}
&& S_{matter}^{g.f.} =
\int _{{\it M}_4} d^4 x \sqrt{g} \  \Tr
\Bigl (\kappa^{\m\n} D_{[\m} \Psi_{\n]}
+ \bar\Phi D^\m D_\m \Phi 
+ \bar\eta D^{\m}\Psi_{\m} \Bigr)
\eea
As for the Yang--Mills supermultiplet, we 
can perform a mapping of the topological ghost
$\Psi_\m$ and of the topological anti--ghosts
$(\kappa^{\m\n}, \bar\eta)$ on left and right
handed spinors $(\psi^\alpha, \bar\psi_{\dot\alpha})$
respectively. 
With this change of variables, we now recognize that the action 
(\ref{HMn=1}) is the $N=1,D=4$ chiral multiplet action
\bea\label{SYMn=1t-chi}
S_{SYM} &=&
\int _{{\it M}_4} d^4x \sqrt{g} \  \Tr
\Bigl (
 \bar\Phi D^\mu D_\mu \Phi  
+
\bar\psi \gamma^\mu D_\mu \psi \Bigr )
\eea
The sum of both  
actions (\ref{BVYM}) and (\ref{BVHM}), 
when the suitable gauge fixing conditions (\ref{Z}) and (\ref{Z'})
are enforced,  
corresponds to the twisted $N=2$ Super Yang--Mills 
action, with notations that are adapted to a K\"ahler manifold
\footnote{In order to recover the Yukawa couplings and 
the quartic term in the potential
$[\bar\Phi,\Phi]^2$ typical of the $N=2$ SYM one should slightly
modify the gauge--fixing fermion $Z'$ in (\ref{Z'}),
but this doesn't change the results on the topological observables.}.

However, the BRST algebra we discussed so far is 
mapped only to a $N=1$ subsector of the $N=2$ supersymmetry.
The complete $N=2$ superalgebra on a K\"ahler manifold has been discussed in
\cite{Park, Witten1, Dijkgraaf1}. Let us briefly display how these results can 
be recovered in our model.
In our construction, we can interchange the role of  holomorphic 
and antiholomorphic coordinates. 
We can thus consider another operator $\bar \Delta$ 
\begin{eqnarray}
    \label{qbar}\let\hw=\hidewidth\matrix
    {
\Delta A _m&= &\Psi  _m   &&  \bar \Delta A _m&= &0 
\cr
\Delta A _\m&= &0 && \bar \Delta A _\m&= & \Psi  _\m 
\cr
\Delta \Psi _m&= & 0    &&  \bar \Delta \Psi _m&= &  D_m \Phi    
\cr
\Delta \Psi _\m&= &   D_\m \Phi     &&  \bar \Delta \Psi _\m&= &   0
\cr
  \Delta \Phi &= &0 &&  \bar  \Delta \Phi &= &0
  }
   \end{eqnarray}
where with $\Delta, \bar \Delta$ we indicate the equivariant
BRST operators, with the ghost field $c$ associated to the gauge symmetry 
set to zero. One has:
     \begin{eqnarray}
    \label{s-tot}\let\hw=\hidewidth\matrix
    {
(\Delta+\bar \Delta) A _m&= &\Psi  _m 
\cr
(\Delta+\bar \Delta)  A _\m& = & \Psi  _\m 
\cr
(\Delta+\bar \Delta)  \Psi _m&= & D_m \Phi    
\cr
(\Delta+\bar \Delta)  \Psi _\m&= &   D_\m \Phi  
\cr
  (\Delta+\bar \Delta)  \Phi &= &0
  }
   \end{eqnarray}
Thus, $\Delta^2=\bar \Delta^2=0$, and $(\Delta+\bar \Delta)^2=\{\Delta,\bar \Delta\}=\delta_\Phi$,  where 
$\delta_\Phi$ is a gauge transformation with parameter $\Phi$.
The operator $(\Delta+\bar \Delta)$ is the topological BRST symmetry operator (for $c=0$), corresponding
to the twisted $N=2$ supersymmetry. 
The classical action which is invariant under the symmetry (\ref{s-tot})
is the (real) BF action plus a ``cosmological term'' $\Tr (B\w B)$,
with $B$ the complete real two--form. This field transform as
$(\Delta + \bar\Delta) B = D\Psi$.
In a space where one cannot consistently separate holomorphic 
and antiholomorphic components of forms, the only admissible operation is     
$(\Delta+\bar \Delta)$, which is Lorentz invariant. 
Then, to close the BRST symmetry, 
and get  $\delta^2=(Q + \bar Q)^2=0$, one must redefine the BRST transformation of the 
Faddeev--Popov ghost $c$, as follows:
 \bea\label{ghost}
Qc= -\demi[c,c] \to  Qc= - \Phi-\demi[c,c]
\eea 
In this way one recover the complete symmetry of the Topological
Yang--Mills action (\ref{fond}).

\section{Six dimensions: Calabi--Yau three--fold}

On a Calabi--Yau three--fold $CY_3$ we can use the holomorphic 
closed $(3,0)$--form $\Omega_{3,0}$ and  define 
$B_{3,1}= \Omega_{3,0} \w B_{0,1}$. The classical action (\ref{hBF})
then becomes
\be
I_{cl} (A,B_{0,1})=\int _{{\it M}_6} \Omega_{3,0} \w  
\Tr (B_{0,1} \w F )
\label{three-f}
\ee
The BRST symmetry corresponding to the action (\ref{three-f}) 
is
\bea
 QA_m &=& \Psi_m +D_m c\ \nn\\
 QA_\m  &=& D_\m c \ , \nn \\
  QB_{\m}  &=& D_{\m}\chi - [c,B_{\m}]   \ , \nn \\ 
  Q\chi &=& - [c,\chi] \ , \nn\\
  Qc  &=&    - \demi[c,c] \ , \nn \\
\label{6-sym}
\eea
The invariance of the action (\ref{three-f}) 
under the transformation of the $B$ field
is guaranteed by part of the Bianchi identity
\be
D_{[\m}F_{\n\bar l]}=0
\label{c-bi}
\ee
and the fact that $\Omega$ is closed. Notice also that 
the action (\ref{three-f})
is invariant under the {\it complexified} gauge group $GL(N,\complex)$.
The BRST symmetry (\ref{6-sym}) follows from the
following Batalin--Vilkoviski action 
\bea
S &=& \int _{{\it M}_6} \sqrt{g} \ d^6x \   
\Tr \Bigl(\epsilon^{\m\p\q } B_{\m} F_{\p\q}
 + ^*B^{\m } ( D_\m \chi - [c,B_{\m }]) \nn\\
&& +
 ^*A^m (\Psi _m +D_\m c)
 +
 ^*A^\m D_\m c  -^* \Psi ^m[c, \Psi _m]
 \nn\\\
&&+^*c ( -\demi [c,c])
\Bigr) \ \ ,
\label{bv-6}
\eea
where we have normalized the $(3,0)$--form $\Omega$ such that
$\Omega\w \bar\Omega$ is the volume form.

Let us now proceed to the quantization of the model: this can be performed
in different ways, which lead us to the study of different sets of observables.
If one chooses to quantize the theory around the perturbative 
vacuum corresponding to holomorphic flat connections, the 
corresponding observables   
will depend on the {\it complex} structure $\Omega$ of the manifold, 
as usually happens
in type B topological string theories. In fact, in this case the holomorphic
BF model has a deep relationship with the holomorphic Chern--Simons theory,
which can be regarded as an effective action for D5 branes in type B topological string \cite{WittenCS}. This relationship 
should be a generalization of that between BF and Chern Simons theories in 
three (real) dimensions \cite{CottaBF}, and it deserves further
investigations.

If instead one quantize the theory around a non-perturbative vacuum 
corresponding to a stable holomorphic vector bundle, one can show that
the BF model correspond to the twisted version of a supersymmetric
Yang--Mills theory! In this case the observables are dependent on
the {\it K\"ahler} data of the manifold, as happens in type A topological 
string. In fact a direct relationship between a twisted $U(1)$
maximally supersymmetric action and the topological vertex has been shown in
\cite{iqbal}.

The study of the holomorphic BF model in the abelian case
can be then useful to clarify the issue
of S--duality in topological strings pointed out in 
\cite{neitzke,ooguri,kapustin}, and the relationship between 
Gromov--Witten and Donaldson--Thomas invariants discussed in \cite{panda}.
Let us now show the details of the two quantizations.

\subsection{Perturbative quantization and B model} 

In this case, 
the symmetries are treated as ordinary gauge
symmetries and fixed with transversality conditions on the $A_\m$ and
$B_\m$ fields:
\bea
D^\m A_\m &=& 0 \nn \\
D^\m B_\m &=& 0
\label{pert-gf}
\eea
The BV fermion corresponding to these conditions
is
\be
Z = \bar\chi D^\m B_\m + \bar c D^\m A_\m
\label{Z-pert}
\ee
Once the gauge fixing conditions (\ref{pert-gf}) are enforced,
one has a well defined mixed propagator between the $A$ and $B$
field, 
and can use it to evaluate the path integral in a perturbative
expansion. The partition function of this model in the semiclassical limit
should be related to the Ray--Singer holomorphic torsion
\cite{RS2} similarly to what happens for the holomorphic Chern--Simons
theory analysed in \cite{DT,Frenkel}. 
The higher order terms in the perturbative expansion should be
related to other manifold invariants. It would be interesting
to study these invariants along the lines of the perturbative
analysis of three--dimensional Chern--Simons theory
\cite{AS1}.

\subsection{Non--perturbative quantization and 
A model}
 
 The shift symmetry on the 
$(0,1)$ part of the connection gives rise to three degrees of freedom,
while the symmetry on the $B$ field to one. These are collected
into the ghost fields $(\Psi_m,\chi)$ respectively.
In the non--perturbative case, the gauge fixing conditions are chosen 
as follows
\bea
F_{mn}&=&- \frac{4}{3} \epsilon_{mnp} B^{p}    \nn \\
J^{\m n}  F_{\m n} &=& 0 
\label{gf-6} 
\eea  
and amount to three conditions for the first line and one for the second.
The reason for the particular choice of the coefficient in the
first equation of (\ref{gf-6}) will be evident shortly.
Notice that the second equation in (\ref{gf-6}) 
reduces the complex gauge group $GL(N,\complex)$ to
the unitary group $U(N)$, and as such can be considered as a partial
gauge--fixing for the complex gauge symmetry of the classical action
(\ref{three-f}). This has to be completed with a further gauge--fixing
for the unitary group, as for example the ordinary Landau gauge
$\partial^\mu A_\mu =0$. We will discuss this issue in more detail
in Sect.5.1.
The BV fermion corresponding to (\ref{gf-6}) is given by
\be
Z= \bar \chi^{\m\n}(F_{\m\n} + \frac{4}{3}\epsilon _{m n p} B^{p}  )
+   \bar \eta (2 i J^{\m n}  F_{\m n} - h) \ \ ,
\label{Z-6}
\ee
where $(\bar\chi^{\m\n}, \bar\eta)$ are the antighosts associated
to the gauge--fixing conditions (\ref{gf-6}), whose BV action
is given by
\be
S_{aux}= \int_{M_6} \sqrt{g} \ d^6x \
\Tr (^*\bar\chi_{\m\n}h^{\m\n} + 
^*\bar \eta h
+^*\bar  c b) 
\label{bv6-aux}
\ee
Eliminating the anti--fields by means of
(\ref{bfZ}) and 
implementing the gauge--fixing conditions (\ref{gf-6})
by integration on the Lagrangian multipliers, we get
from (\ref{bv-6}) and (\ref{bv6-aux})
\bea
S^{g.f.}&=&  \int _{{\it M}_6} d^6x \sqrt{g} \  
\Tr \Bigl(-\frac{3}{2} F^{\m\n} F_{\m\n} + |J^{m\n}F_{m\n}|^2) \nn\\
&& +\bar\chi^{\m\n} D_{[\m}\Psi_{\n]}
  + 2i\bar\eta J^{\m n}D_{\m}\Psi_{n}
  +\frac{4}{3} \epsilon_{\m\n\p} \bar\chi^{\m\n} D^{\p}\chi \Bigr) \ \ .
\label{gfa-6}
\eea 
By using the identity \cite{tian}
\bea
&&-\frac{1}{4} \Tr (F \w *F) + J \w \Tr (F\w F) = \nn \\
&& \Tr \Big(-\frac{3}{2} F^{\m\n}F_{\m\n} + |J^{\m n}F_{\m n}|^2\Big)
\label{id-6}
\eea
we can recognize in the first line of (\ref{gfa-6})
the bosonic part of the N=1 D=6 SYM action, modulo the topological
density $J\w \Tr(F\w F)$, where $J$ is the K\"ahler two--form.
Concerning the fermionic part, we can make use of
the mapping between chiral fermions and complex forms 
$S_{\pm}\otimes \complex \sim \Omega^{^{odd}_{even}}$
to map the topological ghosts $(\Psi_m,\chi)$
into the right--handed spinor $\bar\lambda$ and the topological 
antighosts into the left--handed spinor $\lambda$.
More explicitly, we can use the covariantly constant 
spinor $\zeta$ of the Calabi--Yau three--fold
to perform the mapping
\bea
\Psi_m &\to& \bar\lambda \Gamma_m \zeta \nn \\
\chi &\to& \bar\lambda \zeta \nn \\
\bar\chi^{\m\n} &\to& \zeta \Gamma^{\m\n} \lambda \nn \\
\bar\eta &\to& \epsilon_{\m\n\p}\zeta \Gamma^{\m}\Gamma^{\n}\Gamma^{\p}\lambda
\label{twist6}
\eea
In this way, one can recognize in (\ref{gfa-6}) the twisted version
of the $N=1$ $D=6$ Super Yang--Mills action.
In order to reproduce the $U(1)$ twisted maximally supersymmetric action 
discussed in \cite{iqbal}, one has to add to the classical action
(\ref{three-f}) the higher Chern class $F \w F \w F$ and
couple this theory to an
hypermultiplet, with a procedure similar to that discussed in the four 
dimensional case. 
As in Sect.3.3, one has to consider the quantization of a 
$(0,2)$--form $B_{0,2}$. The corresponding BRST complex
is the same as in Table (\ref{pyr}), but now with six--dimensional fields
($\m,\n=\bar 1,\bar 2, \bar 3$).
It is straightforward to realize that the fields appearing
in the Table (\ref{pyr}) together with the multiplet 
discussed in this subsection give rise exactly to the spectrum
of the twisted maximally supersymmetric Yang--Mills discussed in
\cite{iqbal}.  
An alternative and more economical way 
would be to proceed from the dimensional reduction of the eight-dimensional 
model that we are going to discuss in the next section. 
Notice that, as discussed in Sect.3.3, the coupling to the hypermultiplet 
does not change
the classical action (\ref{three-f}). Moreover, the higher Chern class 
$(F)^3$ is only a boundary term which does not affect the propagation of the
$A$ and $B$ fields. Thus the relationship with the perturbatively quantized
model of the previous subsection still holds.

\section{Eight dimensions}

\subsection{Calabi--Yau four--fold}
 
On a Calabi--Yau four--fold we can write the following
generalization of the action (\ref{BF})
\be
I_{cl} (A,B_{0,2})=\int _{{\it M}_8} \Omega_{4,0} \w  
\Tr (B_{0,2}\w F_{0,2})
\label{four-f}
\ee
Here  $\Omega_{4,0}$ is the holomorphic covariantly {\it closed} (4,0)-form.
This, together with part of the Bianchi identity, ensures the invariance of 
the classical action (\ref{four-f}) analogously to the CY$_3$
case of the previous section. Also, as in the previous section,
we normalize $\Omega$ such that $\Omega\w\bar\Omega$ is the volume element on $M_8$.  
The action (\ref{four-f}) displays the symmetry 
\bea
 QA_M &=& \Psi_M +D_M c\ \nn\\
 QA_\M  &=& D_\M c \ , \nn \\
  QB_{\M\N}  &=& D_{[\M}\chi_{\N]} - [c,B_{\M\N}] 
  -\frac{1}{4}\epsilon_{\M\N\P\Q} [^*B^{\P\Q},\phi]
  \ , \nn \\ 
  Q\chi_\N &=& D_{\N}\phi - [c,\chi_{\N}] \ , \nn\\
  Qc  &=&
    - \demi[c,c] \ , \nn \\
  Q\phi  &=& -[c,\phi]
\label{8-sym}
\eea
Notice that $c$ is the complexified Faddeev--Popov ghost.  
The BV action corresponding to (\ref{four-f}) is given by 
\bea
S &=& \int _{{\it M}_8} \sqrt{g} \ d^8x \
\Tr \Bigl(\epsilon^{\M\N\P\Q} B_{\M\N} F_{\P\Q} \nn \\
&& + ^*B^{\M\N} (D_{[\M}\chi_{\N]} - [c,B_{\M\N}] - 
\frac{1}{4}\epsilon_{\M\N\P\Q} [^*B^{\P\Q},\phi]) \nn\\
&& +
 ^*A^M (\Psi _M +D_\M c)
 +
 ^*A^\M D_\M c + ^*\chi^\N ( D_{\N}\phi - [c,\chi_{\N}]) 
 \nn\\
&&
+ ^*\bar\chi_{\M\N}h^{\M\N} + 
^*\bar \chi h
+^*\bar  c b
+^*\bar\phi 
\bar\eta \nn \\
&&+^*c ( -\demi [c,c])- ^*\phi[c,\phi]
\Bigr)
\label{four-f-gf}
\eea
The action (\ref{four-f}) only define the propagation of  part of the 
gauge field, as in the case studied in section 2.  
It can be gauge--fixed in by imposing  six complex
conditions for $B_{\M\N}$ 
\bea
B_{\M\N}^+&=&0 \ , \nn\\
B_{\M\N}^-&=&F_{\M\N}^- \ , 
\label{gf-8}
\eea 
and a gauge--fixing for $\chi_\M$
\be
D^\M\chi_\M = 0.
\label{chi-gf}
\ee
The projection on self-dual or  anti-self-dual part $B^\pm_{0,2}$
of the $(0,2)$--forms can be done by using the anti-holomorphic $(0,4)$ form. 
The conditions (\ref{gf-8}) can be enforced by using the BRST doublets of 
complex antighosts and Lagrangian
multipliers $(\bar\chi^{\M\N}, h^{\M\N})$ and $(\bar\phi,\bar \chi)$ 
respectively.
Then, as a generalization of  \cite{bakasi}, we complete   the  above   six complex
conditions for $B_{\M\N}$  by the following complex condition :
\bea\label{complex}
D^{\M ~{\bf c}} A_\M=0
\eea
The real part of (\ref{complex}) is the ordinary
Landau gauge condition. The imaginary part 
gives instead a condition analogous to the second line of 
(\ref{gf-6}):
 \bea\label{complex1}
{\it Im}D^{\M} A_\M &=&0 \   \Rightarrow \ \ J^{M\N} F_{M\N}=0  \\
{\it Re}D^{\M} A_\M &=&0 \  \Rightarrow \ \  \partial^\mu A_\mu=0
\nn
\eea
The gauge--fixing fermion corresponding to the gauge conditions
(\ref{gf-8}), (\ref{complex}) and (\ref{complex1}) is
\bea
Z &=&  \bar\chi^{\M\N^+}B_{\M\N}  + \bar\chi^{\M\N^-}(B_{\M\N} - 
2F_{\M\N}) + \bar\phi D^\M\chi_\M
\nn\\
&&+\bar \chi  ( i J^{M\N}F_{M\N}  +\demi h)
+\bar c  ( \partial^{\mu} A_\mu +\demi b)
\label{Z-8}
\eea
By using the BV equation (\ref{bfZ}) and enforcing the
gauge conditions (\ref{gf-8}), (\ref{complex}) and (\ref{complex1})
by integration on the
Lagrangian multipliers, we get the wanted action, as a twisted form of the 
$D=8 $ supersymmetric Yang--Mills action. Its gauge invariant part is :
\bea
S^{g.f.}&=&  \int _{{\it M}_8} d^8x \sqrt{g} \
\Tr \Bigl(2F^{\M\N-} F_{\M\N}^- \nn +\demi |J^{M\N}F_{M\N}|^2) 
+ \bar\phi D^\M D_\M \phi \nn\\
&& +\bar\chi^{\M\N-} D_{[\M}\chi_{\N]}
  +\bar\chi^{\M\N+} D_{[\M}\Psi_{\N]}
  +\bar\chi D^M\Psi_M
+ \bar\eta D^\M\chi_\M \Bigr)
\label{gfa-8}
\eea 
As for the previous cases, we do not display in (\ref{gfa-8}) 
the gauge dependent part of the action.
By using the identity
\be
\int _{{\it M}_8} d^8x \sqrt{g} \
\Tr (2 F^{\M\N-} F_{\M\N}^- + \demi |J^{M\N}F_{M\N}|^2) + S_0 = 
\frac{1}{4} \int _{{\it M}_8} d^8x \sqrt{g} \
\Tr(F^{\mu\nu} F_{\mu\nu}) 
\label{id}
\ee
where $S_0=\int_{M_8}\Omega\w \Tr(F_{0,2}\w F_{0,2})$ is a surface
term \cite{bakasi}, we can recognize the first line of the action
(\ref{gfa-8}) as the bosonic part of the $D=8$ SYM action.
Concerning the fermionic part, one can use the identification
of fermions with forms $S_{\pm} \sim \Omega^{_{odd}^{even},0}$,
which for $D=8$ reads 
\bea
S_- &\sim& \Omega^{1,0} \oplus \Omega^{3,0} \ , \label{s1}\\
S_+ &\sim& \Omega^{0,0} \oplus \Omega^{2,0} \oplus \Omega^{4,0} \ 
\label{s2}
\eea
to identify the topological ghosts $(\Psi_M,\chi_\M)$ with
the right--handed projection of the Majorana spinor 
$\bar\lambda^{\dot a}$, $\dot a=1,\ldots,8$, and 
the topological anti--ghosts
$(\bar\chi, \bar\chi^{\M\N},\bar\eta)$ with the left--handed
projection $\lambda_{a}$. 
Notice in fact that on a Calabi--Yau four--fold one can use the
holomorphic four--form $\Omega$ to identify the $(3,0)$--form 
appearing on the r.h.s. of (\ref{s1}) with the
the field $\chi_\M$ \cite{lawson}.
Analogously, one can identify the scalar $\bar\eta$ with the 
$(4,0)$--form appearing in the r.h.s. of (\ref{s2}).
More explicitly, one can use the two left--handed
covariantly constant spinors $(\zeta_1, \zeta_2)$
of the Calabi--Yau four--fold to identify
\bea
\chi^M &=& \epsilon^{MNPQ} \zeta_1\Gamma_N\Gamma_P\Gamma_Q\bar \lambda  \nn\\
\bar\chi_{MN}^ - &=&\zeta_1 \Gamma_{MN}^-  \lambda \nn\\
\bar\chi &=&  \zeta_1 \lambda 
\label{8-twist}
\eea
and
\bea
\Psi_M &=&  \zeta_2 \Gamma_M \bar \lambda \nn\\
\bar\chi_{MN}^ + &=&  \zeta_2 \Gamma_{MN}^+ \lambda \nn\\
\bar\eta &=& \epsilon^{MPNQ} \zeta_2 \Gamma_M\Gamma_N\Gamma_P\Gamma_Q \lambda
\label{8-twist2}
\eea

{\bf moduli space and $Spin(7)$ theory}: the moduli space probed by the above TQFT is
a holomorphic  $(0,2)$--form $\bar D B_{02}=0$
and
\bea
F^+_{\M\N} &=& 0 \label{cplx-sd}\\
D^{\M{\bf c}} A_\M &=& 0
\label{cplx-moduli}
\eea
Notice that the classical action (\ref{four-f}) is invariant
under the group of {\it complex} gauge transformations
$GL(N,\complex)$. The moduli space described by (\ref{cplx-sd}) and 
(\ref{cplx-moduli})
with complex gauge group $GL(N,\complex)$ should be equivalent to that
described by (\ref{cplx-sd}) and 
(\ref{complex1}) with the unitary group $U(N)$. 
This last moduli space is directly related with that explored by
a $Spin(7)$ invariant topological action \cite{DT}.
In fact, by using (\ref{complex1}) one realizes that the imaginary part 
of (\ref{cplx-moduli}) together with (\ref{cplx-sd}) amount 
to seven real conditions, which 
fit in the ${\bf 7}$ part of the $Spin(7)$
decomposition of the (real) two--forms ${\bf 28} = {\bf 7} \oplus {\bf 21}$.
The real part of (\ref{cplx-moduli}) is the ordinary transversality
condition for the unitary gauge group.
Then the theory defined by the action (\ref{four-f})
should be equivalent to that defined by the $Spin(7)$--invariant action
\be
I_{\Psi-BF} = \int_{M_8} \Psi \w \Tr(B\w F)
\label{Psi-BF}
\ee
where $\Psi$ is the real $Spin(7)$--invariant Cayley four--form
and $B,F$ are real two--forms.
The mapping can be done by identifying the fundamental 
representation of the $SU(4)$ group with the real spinor representation 
of $Spin(7)$.

 \subsection{Seven dimensions: from CY$_4$ to CY$_3\times S^1$}
 
In \cite{bakasi}, the case of writing a 
BRSTQFT for a $G_2$ manifold was directly done by  starting from the topological 
action 
 \bea
 \int _{\it M_7} d^7x   c^{ijk} D_i\varphi F_{jk}
 \eea
  where $c^{ijk} $ stand for the $G_2$-invariant tensor made of octonionic 
structure coefficients, and $\varphi $ is  a Higgs field. The BRST quantization of this topological term yields 
a twisted version of the dimensional reduction to seven dimensions of the 
D=8 super Yang--Mills action on a manifold with $Spin(7)$ or $SU(4)$ holonomy. Basically, the topological gauge functions 
are the generalization of Bogolmony equations in 7 dimensions, 
as shown in \cite{bakasi}.
  
$G_2$ manifolds of the kind  
$\Sigma_6\times S_1$, where $\Sigma_6$ is a Calabi--Yau 3-fold, 
are of special interest both for mathematical \cite{Hitchin} and
physical \cite{Dijgk} applications. A topological theory for such manifolds
can be obtained by considering the dimensional reduction of the
model discussed in the previous section, for Calabi--Yau four--fold, 
although we will shortly give the classical 7-dimensional action that one
can directly quantize on such manifolds.

Starting from the ATQFT for $CY_4$,  one can set the fourth component of the gauge connection to
$A_{\bar 4} = A_7 - i L$, where $7$ is the direction along the circle
$S_1$ and $L$ a real scalar field.  (L will shortly have a special interpretation in seven dimensions.) Moreover, the dimensional reduction
impose to set 
$i(\partial_4 -\partial_{\bar 4}) A_{\bar M} = \partial_8 A_{\bar M} = 0$
for any $\bar M = \bar 1, \ldots, \bar 4$. Then one gets from
the classical action (\ref{four-f}) the following action 
\be
S = \int _{{\it \Sigma}_6\times S^1} \sqrt{g} \ d^7x \   
\Tr \Big[
2\epsilon^{\m\n\p} (B_{\m\n} F_{\p\bar 4} + B_{\m}F_{\n\p})
\Big]
\label{7d-cl}
\ee
with $F_{\p\bar 4} = F_{\p 7} - i D_\p L$ and $B_\m = B_{\m \bar 4}$. 
In this section 
we define 
the complex three--form $\Omega_{0,3}$ on $\Sigma_6$ starting
from the (normalized) complex four--form in eight dimensions as
$\epsilon_{\m\n\p}=\epsilon_{\m\n\p\bar 4}$.  
We thus 
consider the following classical action:  
  \be
I_{cl} (A,B_{0,2})=\int _{{\it M}_7} \Omega_{3,0}  \w  
\Tr (B \w F )
\label{omega-7}
\ee
Here the coordinates on the manifold are $z^m,z^\m$ for $\Sigma_6$  and the periodic real coordinate $x^7$ for the circle. The only components of the 2-form that have a relevant  propagation   are  $B_{\m\n}, B_{\m7}$.

The covariant quantization of $B_{\m\n}, B_\m(=B_{\m7})$ and $A$ requires 
seven topological antighosts
$(\kappa^{\m\n}, \kappa^\m, \kappa)$. Modulo the ordinary gauge symmetry,
we have indeed seven freedoms for the classical action (\ref{7d-cl})
associated to the topological ghosts 
$(\chi_{\m} , \chi)$ for the fields $(B_{\m\n}, B_\m)$
and the ghost $\Psi_m$ for the field $A_m$. The relevant invariance is $B_{\m\n} \sim  B_{\m\n} +D_{[\m}\chi_{n]}$,
$B_{\m } \sim  B_{\m } +D_{ m}\chi $
 $A_m \sim  A_m+\Psi_m$, $A_\m \sim  A_\m $, $A_7 \sim  A_7, $ modulo ordinary gauge transformations.  
Moreover, in the BRST complex for the field $B$ it will appear
also three commuting scalar ghost of ghosts ($\Phi$, $L$, $\bar \Phi$),
for the "gauge symmetries" of the topological ghosts and antighosts of $B_2$. All these fields are conveniently displayed 
as elements of the following pyramidal diagram:
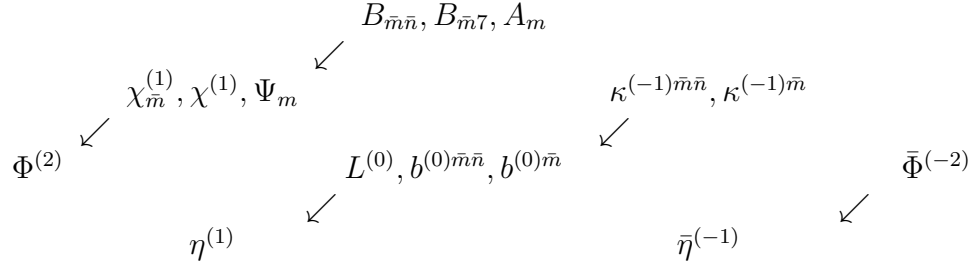
\begin{eqnarray}\nn
\let\hw=\hidewidth
\matrix{ && &&   &&    &&  B_{\m\n} , B_{\m7}, A_m     
\cr   &&    &&    &&         &   \hw\swarrow\hw  
\cr   && &&   &&     \chi^{(1)}_{\m},  \chi^{(1)} ,\Psi_m &&    &&        
\kappa^{(-1)\m\n}, \kappa^{(-1)\m }  
\cr  && &&   &  \hw\swarrow\hw &&    &&   \hw    \ \ \ \ \ \ \    \swarrow\hw  
\cr  && &&   \Phi^{(2)}  &&    &&      \hw L^{(0)}, b^{(0)\m\n}  ,b^{(0)\m}
\hw &&  &&   \bar\Phi^{(-2)} 
\cr  
&& &     &&&& \hw\hw  \hw\swarrow\hw&&&&  \swarrow\cr  && &&   &&   \hw   \eta^{(1)}
\hw &&  &&   \hw \bar\eta^{(-1)} \hw &&      &&      \cr   }%
\end{eqnarray}%
\vskip 0.5 cm
\noindent
From the point of view of the dimensional reduction of the CY$_4$ theory, it   is interesting to observe 
that the medium ghost of ghost $L$ of the $B$ field
can be identified with the component $A_8$ 
of the eight dimensional gauge field.  
The BRST transformations of the fields can be read from the following   Batalin--Vilkoviski action :
\bea
S &=& \int _{{\it \Sigma}_6\times S^1} \sqrt{g} \ d^7x \   
\Tr \Big[
2\epsilon^{\m\n\p} (B_{\m\n} F_{\p\bar 4} + B_{\m}F_{\n\p})
\nn \\
&&
+ ^*B^{\m\n} ( D_{[\m} \chi_{\n]} - [c,B_{\m\n}]
+ \demi \epsilon_{\m\n\p} [^*B^{\p},\Phi ]) 
\nn \\
&& 
+ ^*B^{\m} (2 D_{[\m}\chi_{7]} - 2 [c,B_{\m}] 
+ \demi \epsilon_{\m\n\p}[^*B^{\n\p}, \Phi]) 
\nn\\
&& 
+ ^*\chi^{\m} ( D_{\m} \Phi - [c,\chi_{\m}])
+ ^*\chi (D_{\bar 4} \Phi - [c, \chi] )
\nn\\
&& 
+ ^*A^m (\Psi _m +D_\m c)
+ ^*A^\m D_\m c   +  ^*A^7D_7 c  -^* \Psi ^m[c, \Psi _m]
\nn\\
&&
+ ^*c ( -\demi [c,c])
+ ^*\bar \Phi  \bar \eta +^*L\eta
 \nn\\
&&
+^*\kappa_{\m\n} b^{\m\n}+ 2^*\kappa_{\m} b^{\m}
+^*\kappa b
\Bigr] \ \  
\label{bv-7}
\eea
This action is actually the dimensional 
reduction of  action (\ref{four-f-gf}). However, the interpretation of the ghost of ghost system is quite different.
Notice that the vectorial part of the
BRST transformation for the field $B_\m$
is
\be
\delta B_{\m}=D_\m \chi - D_{\bar 4}\chi_\m = D_\m\chi - D_7 \chi_\m
\label{qbm}
\ee
The variation of the action due to the first term in (\ref{qbm}) simply
vanishes after integration by parts because of the identity 
(\ref{c-bi}). In the context of the dimensional reduction,
this identity can be read as the $\M=\bar 4$ component
of the eight--dimensional one $\epsilon^{\M\N\P\Q}D_{\M}F_{\N\Q}=0$.   
The variation associated to the last term in (\ref{qbm}), 
together with that coming from the variation of
the field $B_{\m\n}$, 
$QB_{\m\n} = D_{[\m}\chi_{\n]}$,
gives after integration by parts the
other three components $\M=\bar 1,\bar 2,\bar 3$ 
of the above eight--dimen\-sional identity.
The topological freedom of the classical action (\ref{7d-cl}) can be
fixed by choosing seven independent gauge functions. 
The first six can be directly obtained from  
the dimensional reduction of (\ref{gf-8})
\bea
B_{\m\n} &=& \demi (F_{\m\n} - \epsilon_{\m\n\p}F^{\p\bar 4}) \nn \\
B_{\m} &=& \demi (F_{\n\bar 4} - \demi \epsilon_{\n\p\q} F^{\p\q})\nn \\
&=& \demi (F_{\n 7} - i D_\n L  - \demi \epsilon_{\n\p\q} F^{\p\q})
\label{gf-7} 
\eea  
while the seventh one corresponds to the imaginary part of the complex 
gauge--fixing condition (\ref{complex})
\be
\demi J^{m\n}F_{m\n} + D_7 L = 0
\label{real-7}
\ee
Notice in fact that the reduction on a manifold of the kind 
$\Sigma_6\times S^1$ breaks the complex group of gauge invariance
of the eight--dimensional action (\ref{four-f}) to the 
unitary group. The residual gauge invariance under this group
can be fixed by the ordinary transversality condition 
$\partial^\m A_\m + \partial^m A_m + \partial^7 A_7 =0$.
The gauge--fixing fermion corresponding to the conditions (\ref{gf-7}) 
and (\ref{real-7}) is
\bea
Z&=&
\kappa^{\m\n}\left[B_{\m\n} - \demi \left(
F_{\m\n} - \epsilon_{\m\n\p}(F^{\p 7} + i D^\p L)\right)\right]
\nn\\
&& + \kappa^\m \left(B_\m - \frac{3}{2} (F_{\n 7} - i D_\n L - 
\demi \epsilon_{\n\p\q}F^{\p\q})\right)
\nn\\
&& + \kappa \Bigl( 2 (J^{\m n}  F_{\m n} + D_7 L) - b\Bigr) \nn\\
&& + \bar\Phi (D^\m \chi_\m + D^7 \chi) 
\label{Z-7}
\eea
The bosonic part of the gauge--fixed action, which can be obtained as usual
by using the BV equation (\ref{bfZ}) and enforcing the gauge conditions
(\ref{gf-7}) and (\ref{real-7}) by integration on the Lagrangian multipliers,
reads
\bea
S^{g.f.}&=&  \int _{{\it M}_7} d^7x \sqrt{g} \  
\Tr \Bigl(-\frac{3}{2} F^{\m\n} F_{\m\n} + |J^{m\n}F_{m\n}|^2 
- \demi F^{\m7} F_{\m7}
\nn\\
&&
+\bar \Phi (D^\m D_\m + D^7 D^7) \Phi + 
L (D^\m D_\m + D^7 D^7) L \Bigr)
\label{gfa-7}
\eea 
and can be identified with the bosonic part of the seven dimensional 
Super Yang--Mills on $M_7 = \Sigma_6 \times S^1$.

Concerning
the fermionic sector,  
we have 8 topological ghosts, $\Psi_\m, \chi_{\m}, \chi$ and $\eta$, 
and 8 topological antighosts 
$\kappa^{\m\n},\kappa^\m, \kappa$ and $\bar \eta$. 
The mapping with spinors can be obtained from the dimensional reduction
of the eight--dimen\-sional mapping (\ref{8-twist}) and (\ref{8-twist2}).
After the dimensional reduction to $\Sigma_6\times S^1$,
the two covariantly--constant eight--dimensional chiral spinors 
$(\zeta_1,\zeta_2)$ are 
identified with the unique covariantly constant Majorana spinor $\xi$ of
the Calabi--Yau three--fold $\Sigma_6$. On the other side, the eight
dimensional spinor $\lambda$ yields 
two seven dimensional Majorana spinors $(\lambda_1, \lambda_2)$.
This results in the mapping
\bea
&& \Psi_m \to \xi \Gamma_m \lambda_1 \ \ , \ \ 
\eta \to \xi \Gamma_7 \lambda_1 \nn \\
&& \chi^\m \to \epsilon^{mnp} \xi \Gamma_{np} \lambda_1 \ \ , \ \
\chi \to \epsilon^{mnp} \xi \Gamma_{mnp} \lambda_1 
\label{twi-gh}
\eea
for the topological ghosts and 
\bea
&& \kappa_{mn} \to \xi \Gamma_{mn} \lambda_2 \ \ , \ \
\kappa_{m} \to \xi \Gamma_{m 7} \lambda_2 \nn \\
&& \kappa \to \xi \lambda_2 \ \ , \ \ 
\bar\eta \to \epsilon^{mnp}\Gamma_{mnp7}\lambda_2
\label{twi-agh}
\eea
for the topological antighosts. 
By this mapping we can identify the topological action
(\ref{7d-cl}) gauge fixed with the conditions
(\ref{gf-7}) and (\ref{real-7}) as the twisted version 
of $N=2$ seven dimensional Super Yang--Mills\footnote{
As discussed for the four dimensional case, in order to recover 
the Yukawa couplings and 
the quartic term in the potential
$[\bar\Phi,\Phi]^2$ typical of the $N=2$ SYM one should slightly
modify the gauge--fixing fermion (\ref{Z-7}).}. 
The observables of the topological model can be identified
with the dimensional reduction of the eight-dimensional cocycles.

\subsection{Manifolds with $SU(4)$ group structure and self--dual Yang--Mills in eight--dimensions}

On a K\"ahler manifold with a  $SU(4)$ group structure 
 we can choose $B_{4,2}= J\w J \w B_{2,0}^+$,
where $J$ is as usual the K\"ahler $(1,1)$ form.
Here the 2-form $ B_{2,0}^+$ is self dual in the indices $[mn]$.  Self duality is defined 
from  a $SU(4)$ invariant 4-form $\Omega_{4,0}$, which is globally well defined, but not necessarily closed.
It counts for 3 degrees of freedom, according to 
the $SU(4)$ independent decomposition of a 2-form in 8 dimensions:
\be
28=6\oplus \bar 6\oplus 15\oplus 1
\ee
and a further decomposition  $6$ as $6=3 \oplus 3$, using the $\epsilon_{mnpq}$ tensor.

The novelty of this case is that neither    K\"ahler $(1,1)$-form nor $\Omega_{4,0}$ are    necessarily closed. 
Both the forms can also be rewritten in terms of spinors, 
which correspondingly are not covariantly constant with respect to
the usual spin connection, but only with respect to a modified 
connection including torsion terms. 
The corresponding classical action is a generalisation
of that in (\ref{BF}):
\be
I_{k-BF} (A,B_{2,0}^+,B_{0,2}^+)  = \int_{M_8} J\w J \w \Tr (B_{2,0}^+\w F_{0,2})
\label{kBF}
\ee
The symmetries of this action are
\bea
&&Q A_m = \Psi_m \, \quad\quad Q A_\m = D_\m c \nn\\
&&Q B_{mn}^+ = -[c,B_{mn}^+] \ , \quad\quad Q B_{\m\n \ +} = 
\left(D_{[\m}\Psi_{\n]}\right)^+
- [c,B_{\m\n}^+]+[^*B^{mn \ +} ,\Phi] \nn \\
&&Q \Psi_m = 0\ , \quad\quad Q \Psi_\m = D_\n \Phi - [c,\Psi_\n]
 \nn \\
&&Q \Phi = -[c,\Phi] \ , \quad\quad Qc = -\demi[c,c] 
\label{kBF-sym}
\eea
The quantization of this action can be worked out using the BV formalism,
and is very similar to that already discussed in \cite{bakasi}.

We have 8 freedoms for gauge-fixing the system ($B_{2,0}^+, B_{0,2}^+, A_{1,0}, A_{0,1}$). Indeed, 
$\Psi_m$, $\Psi_\m$ and $c$  have respectively 4,4 and  1  components, but  $\Psi_\m$  has a ghost of ghost symmetry with ghost of ghost $\Phi$, so it only counts for 3=4-1 freedoms. 
We can choose  the following 7 gauge--fixing conditions in the gauge covariant sector:
\bea \label{k-BF-gf}
B_{mn}^+ &=& \epsilon_{mnlp\m\n\l\p} F^{\m\n+} J^{l\l} J^{p\p} \\
B_{\m\n^+}&=& 0 \nn \\  
J^{m\n} F_{m\n} &=& 0  \nn
\eea  
plus the ordinary transversality condition for the gauge field, 
$(\partial^m A_m + \partial^\m A_\m=0)$. 
The transformation law of $B_{mn}^+$ implies that a gauge fixing for $\Psi_\m$ must be also done, with a gauge function :
\be
D^\m \Psi_\m=0
\ee
The $Q$-invariant  gauge fixing of the action with these functions  is standard, and reproduces the action as in \cite{bakasi}, that is the twisted form of the 
eight--dimensional Super Yang--Mills theory.
The classical action (\ref{kBF}) can be considered as an eight--dimensional
generalization of the self--dual Yang--Mills in four dimensions,
in particular of its realisation studied in \cite{siegel}.
We should notice that we are at the extreme point of the definition of an ATQFT. The action is the sum of a d-closed term and a Q-exact term.  Thus, there is a ring of observables defined from the cohomology of the BRST operator. However, the classical action is actually completely dependent of the metrics of the manifold, since it depends on both
the K\"ahler form and the complex form at the same time. 

\section{BF theory and Topological M--theory}

Let us consider the following real BF model on a six--manifold $M$
\be
S_{BF,\alpha} = \int_{M} \Phi \w \Tr (BF_A + \frac{\alpha^2}{3} B^3)
\label{bf6}
\ee
where $\Phi$ is a real three--form, $B$ a one form and $A$
a gauge connection with curvature $F_A$. Both $A$ and $B$
are valued in the adjoint representation of an unitary 
gauge group.  
On a Calabi--Yau manifold $X$, the partition function 
of this theory 
\be
Z_{BF,\alpha}(X,\beta) = \int {\cal D}A {\cal D} B 
\ {\rm e}^{-\beta S_{BF,\alpha}}
\label{z-bf}
\ee
can be formally identified with the square of the 
partition function of the holomorphic Chern--Simons
\be
Z_{hCS} (X,l) = \int {\cal D}{\cal A} 
\ {\rm e}^{- l S_{hCS}(\A)}
\label{z-hcs}
\ee
with 
\be
S_{hCS}(\A) = \int \bar\Omega \w \Tr (\A\partial \A + \frac{2}{3} \A^3)
\label{hcs}
\ee 
provided that we identify the real three--form $\Phi$ as the 
imaginary
part
of the holomorphic $(3,0)$--form $\Omega$ of the Calabi--Yau
three--fold and
make the following identifications
\bea
A &=& \frac{1}{2} (\A + \bar\A) \nn \\
B &=& - \frac{i}{2\alpha} (\A - \bar\A)
\label{cv}
\eea
and $\beta = 4l\alpha$.
This is the generalisation to six dimensions of the 
relationship already known in three dimensions \cite{CottaBF}
between the BF theory (and the associated Turaev--Viro invariant \cite{TV} )
and the Chern--Simons theory (and the associated Reshetikhin--Turaev invariant \cite{RT}).

The advantage of the action (\ref{bf6}) is that on some special manifolds
it can be directly related to gravitational theories.
For example, one can lift (\ref{bf6}) to a seven--dimensional manifold as
\be
S_{BF,7} = \int_{N} G \w \Tr (BF_A + \frac{\alpha^2}{3} B^3) \ .
\label{bf7}
\ee 
On $G_2$ manifolds which can be described as the fibration 
of a spin bundle over a three--manifold (or a four--manifold),
the term $(BF + B^3)$ can be identified with the associative three--form
of the Hitchin's action by considering the field $B$ as
the vielbein and the field $A$ as the spin connection (see Sect.6 of \cite{Dijgk}).
This allows us to relate the BF model (\ref{bf7}) to the Hitchin's
action functional.
Notice that 
if $N=X\times S^1$ we can take \footnote{A similar analysis can be done
if we substitute the circle $S^1$ by the real line $\real$ or an interval. In the last case,
boundary terms have to be taken into account.} 
\be
G = {\rm Im}
(\Omega) dt + \demi J\w J\ ,
\label{G2}
\ee
where $t$ is the 
coordinate on $S^1$ and $J$ the K\"ahler
form on X. 
Then by choosing the temporal gauge $A_t=B_t=0$ for the
$A$ and $B$ fields one recover 
the partition function (\ref{z-bf}),
where now $\beta$ is the size of the circle $S^1$.

\vspace{.5cm}

{\bf Acknowledgments}: L.B. is grateful to I.M. Singer    
for interesting discussions.    
A. T. would like to thank U. Bruzzo for useful discussions and all members  
of LPTHE where most of this work has been done.

\end{document}